\documentclass[uplatex, a4paper,12pt]{article} 
\usepackage[utf8]{inputenc}
\usepackage{amsmath}
\usepackage{color}
\usepackage{amssymb, amsfonts, amsthm, amscd}
\usepackage{bm}
\usepackage{float}
\usepackage{caption}
\usepackage[subrefformat=parens]{subcaption}
\usepackage{hyperref}  
\usepackage{graphicx}  

\usepackage{bookmark}            
\hypersetup{
  setpagesize=false,
  bookmarksnumbered=true,
  bookmarksopen=true,
  colorlinks=true,
  linkcolor=blue,
  citecolor=red,
}
\begin{document}

\thispagestyle{empty}
\vspace*{-15mm}

\begin{flushleft}
{\bf OUJ-FTC-24}\\

\end{flushleft}

{\bf }\

\vspace{15mm}

\begin{center}
{\Large\bf
Possible mixing between elementary and bound state fields in the $t\bar{t}$ production excess at the LHC}

\baselineskip 18pt
\vspace{7mm}


Yoshiki Matsuoka

\vspace{5mm}

{\it Nature and Environment, Faculty of Liberal Arts, The Open University of Japan, Chiba 261-8586, Japan\\

}

\end{center}

\vspace{3cm}

\begin{flushleft} 
Email: machia1805@gmail.com  
\end{flushleft}

\vspace{10mm}
\begin{center}
\begin{minipage}{14cm}
\baselineskip 16pt
\noindent

\begin{abstract}
Recent report by CMS Collaboration on the excess of top and anti-top pair production is studied, under  the hypothesis of the coexistence of a toponium $(\eta_t)$ and an additional elementary field $(\Psi)$. We examine the scenario where toponium and an additional field are mixed, and consider the plausible scenarios in that case.

Two scenarios are examined: one is the minimal model with $\Psi$ close to the inert Higgs doublet, and the other is embedded into the two Higgs doublet models (2HDM), where $\Psi$ is one of the two Higgs scalars after transforming the basis. The value of the each coupling constant is restricted by the Multicritical Point Principle (MPP).

Consistency with the data gives constraints on a mixing angle $\theta\ (-45^\circ\le\theta\le45^\circ)$, with which the mass eigenstate $\Psi^\prime$ contributing to the excess is defined by $\Psi^\prime=\Psi\cos \theta + \eta_t\sin \theta$.  The obtained results are $|\theta| \le 13^{\circ}$ for the minimum scenario, and $|\theta| \le 1^{\circ}$ for the second scenario of 2HDM(Type II and Y). We also briefly discuss the comparison with Type I and X.

\end{abstract}

\end{minipage}
\end{center}

\vspace{0.5cm}
\newpage
\section{Introduction}

The Large Hadron Collider (LHC) continues to search for physics beyond the Standard Model (SM)\cite{higgs2,LHC2,LHC3,LHC1}. Recently, the CMS and ATLAS Collaborations reported an excess in the invariant-mass spectrum of top–antitop pairs ($t\bar{t}$) near threshold, based on $pp$ collisions at $\sqrt{s}=13\ \mathrm{TeV}$ with an integrated luminosity of $138\ \mathrm{fb}^{-1}$\cite{topo_1,topo_2,topo_3,topo_4, toponium,topo_5,toponium_atlas}. The excess attains local statistical significance and has attracted considerable attention. According to the CMS analysis, this is a pseudo-scalar toponium bound state with $J^{\mathrm{PC}}=0^{-+}$\cite{toponium_1}. As another possibility, one can hypothesize an unknown (pseudo-)scalar, and the possibility that this state coexists with toponium has also been discussed\cite{toponium}.

In this work, we consider mixing between an additional elementary field $\Psi$, which couples to the top quark through a new Yukawa interaction, and the toponium bound state. As a motivation, \cite{topotopo3} suggests that if only the bound state $\eta_t$ exists within the SM, the stability of the Higgs vacuum may be driven in a more unstable direction, which can in some cases further destabilize the vacuum. Therefore, we also intend to introduce an additional elementary field to address this issue. When the bound state $\eta_t$ mixes with $\Psi$, the mass eigenstates $\Psi^\prime$ and $\eta_t^{\prime}$ are formed. We propose and analyze such mixing scenarios and delineate the phenomenologically viable parameter space, with particular emphasis on the magnitude of the mixing angle.

Renormalization Group control is essential even in the presence of the composite state $\eta_t$. To handle this, we adopt the Bardeen-Hill-Lindner (BHL) framework\cite{topotopo3,topotopo,topotopo2}, in which the composite bound state is treated as an effective elementary field while imposing the compositeness condition $Z_2(\mu=\Lambda)=0$ at a cutoff scale $\Lambda$.

We then develop two complementary realizations. First, in a minimal mixing scenario, the elementary $\Psi$ and the bound state $\eta_t$ mix to form a physical eigenstate $\Psi^\prime$ that can contribute to the excess. The viable parameter region is identified using the Multicritical Point Principle (MPP)\cite{MPPoriginal,MPPoriginal2,MPPoriginal3,kk1,mppeg1,kawai1,mppeg2,me1} as a high-scale organizing principle. Second, we embed this framework into Two-Higgs-Doublet Models (2HDM)\cite{DM2} and analyze the ensuing constraints. We compare the two setups and argue that the minimal mixing scenario both relaxes phenomenological bounds and offers greater predictivity.

The paper is organized as follows. Section \ref{sec2} reviews the MPP and its application to the one-loop effective potential. Section \ref{sec3} presents the model, derives the RGEs, and determines the MPP-consistent parameter space. Section \ref{sec4} extends the framework to the 2HDM and analyzes the associated constraints. Section \ref{sec5} provides a summary and outlook. \ref{firstappendix} lists the one-loop RGEs for the minimal model. \ref{secondappendix} gives the one-loop effective potential for the 2HDM$+$toponium model, and \ref{thirdappendix} presents its one-loop RGEs.

\section{About MPP}\label{sec2}
We briefly summarize the MPP. For the renormalization scale $\mu$, the MPP requires the one-loop Higgs effective potential $V_{\mathrm{eff}}$ to exhibit (approximately) degenerate stationary minima at the electroweak scale $\mu_{\mathrm{EW}}$ and at a much higher scale $\mu_c \gg \mu_{\mathrm{EW}}$:
\begin{align}\label{mppcd}
V_{\mathrm{eff}}(\mu_{\mathrm{EW}}) &= V_{\mathrm{eff}}(\mu_c) \simeq 0,\\
\frac{d V_{\mathrm{eff}}}{d\mu}\Big|_{\mu=\mu_{\mathrm{EW}}} &= \frac{d V_{\mathrm{eff}}}{d\mu}\Big|_{\mu=\mu_c} \simeq 0.
\end{align}
The first relation enforces multi\-criticality (degenerate vacua at widely separated scales), while the second requires approximate stationarity. Since $\mu_{\mathrm{EW}}$ and $\mu_c$ are hierarchically separated and a priori unrelated, realizing degeneracy implies $V_{\mathrm{eff}}(\mu_c) \approx 0$. The MPP has shown notable predictive power, e.g., for the top-quark and Higgs masses\cite{MPPoriginal,MPPoriginal2}, suggesting that unknown high-scale dynamics tune the potential toward multi\-criticality.

\section{Minimal scenario to explain the excess in LHC}\label{sec3}
In this section, we extend as follows: a toponium bound state $\eta_t$ and an elementary field $\Psi$ which is like the inert Higgs doublet\cite{2hdm}. A new Yukawa interaction $y_\Psi$ is introduced, while the couplings of $\Psi$ to fermions other than the top quark are set to zero.

To include the short-distance effects near the threshold around toponium, we follow the Bardeen-Hill-Lindner (BHL) prescription\cite{topotopo3,topotopo,topotopo2}. At a cutoff scale $\Lambda$, an effective four-fermion interaction for top quarks induces a wavefunction renormalization for $\eta_t$, and we impose the boundary condition $Z_2(\Lambda)=0$. This allows us to treat $\eta_t$ as an effective elementary field for $\mu\leq\Lambda$ and provides a consistent low-energy Effective Field Theory (EFT). In this framework we set the vacuum expectation values (VEVs) of $\Psi$ and $\eta_t$ to zero. 

More concretely, we start by introducing a four-fermion interaction among top quarks with coupling $G$,
\begin{eqnarray}
G(\bar{Q}_{3L} t_R)(\bar{t}_R Q_{3L})
\end{eqnarray}
where $Q_{3L}=(t_L,b_L)^T$. By applying a Hubbard--Stratonovich (HS) transformation, we linearize this interaction via an auxiliary scalar field $\tilde{\eta}_t$:
\begin{eqnarray}
\exp\!\left[i\int d^4x\,G(\bar{Q}_{3L} t_R)(\bar{t}_R Q_{3L})\right]
= \int D\tilde{\eta_t}\,D\tilde{\eta_t}^\dagger \,
\exp\!\left[i\int d^4x\left(
-\frac{1}{G}\tilde{\eta_t}^\dagger \tilde{\eta_t}
+\tilde{\eta_t}^\dagger \bar{t}_R Q_{3L}
+\bar{Q}_{3L} t_R \tilde{\eta_t}
\right)\right].
\end{eqnarray}
After an appropriate field normalization, this construction yields an effective Yukawa interaction and a scalar mass term,
\begin{eqnarray}
y_{\eta_t}\,\bar{Q}_{3L} \tilde{\eta_t} t_R + (\mathrm{h.c.}) - M_{\eta_t}^2\,\eta_t^\dagger\eta_t.
\end{eqnarray}
Following the Bardeen--Hill--Lindner (BHL) prescription, we impose compositeness boundary conditions at the cutoff scale $\mu=\Lambda$: the wavefunction renormalization of $\eta_t$ vanishes,
$Z_2(\Lambda)=0$ (so that the kinetic term disappears), and equivalently
\begin{eqnarray}
\frac{1}{y_{\eta_t}(\Lambda)}\to 0 \quad \left(\text{i.e. } y_{\eta_t}(\Lambda)\to\infty\right)
\end{eqnarray}
so that $\eta_t$ does not behave as an elementary degree of freedom at $\Lambda$.
Moreover, we set the quartic self-coupling to
\begin{eqnarray}
\lambda_{\eta_t}(\Lambda)=0
\end{eqnarray}
since it is generated radiatively by top-quark loops below $\Lambda$.
In addition, a mass term may be introduced by hand at the cutoff scale, and other quartic interactions can likewise be induced through top-loop effects. The choice of the cutoff $\Lambda$ will be discussed later.

In addition to $y_{\eta_t}$, we also include interactions involving the new field $\Psi$. The additional Yukawa terms and the additional tree-level scalar potential are given by:

\begin{eqnarray}
-\mathcal{L}_{\mathrm{Yukawa}}=y_{\eta_t}\bar{Q}_{3L}\tilde{\eta_t}t_R+y_{\Psi}\bar{Q}_{3L}\tilde{\Psi}t_R+(\mathrm{h.c.})
\end{eqnarray}
and
\begin{eqnarray}
V(\Psi,H,\eta_t) &=& M_{\Psi}^2(\Psi^\dagger \Psi)+M_{\eta_t}^2(\eta_t^\dagger\eta_t)+M_{\Psi\eta_t}^2(\Psi^\dagger\eta_t+\mathrm{h.c.})\nonumber\\
&+&\frac{\lambda_\Psi}{2}(\Psi^\dagger \Psi)^2+\frac{\lambda_{\eta_t}}{2}(\eta_t^\dagger \eta_t)^2\nonumber\\
&+&\kappa_{1}(H^\dagger H)(\Psi^\dagger \Psi)+\kappa_{2}(H^\dagger H)(\eta_t^\dagger \eta_t)\nonumber\\
&+&\kappa_{3}(\Psi^\dagger \Psi)(\eta_t^\dagger \eta_t)+\kappa_{4}(\Psi^\dagger \eta_t)(\eta_t^\dagger \Psi)\nonumber\\
&+&\frac{\kappa_{5}}{2}((\Psi^\dagger \eta_t)^2+\mathrm{h.c.})
\end{eqnarray}
where the couplings $\kappa_i$ ($i = 2,3,4,5$) and $\lambda_{\eta_t}$ are set to vanish at a cutoff scale, $\kappa_i(\Lambda) = \lambda_{\eta_t}(\Lambda) = 0$, while the Yukawa coupling is taken to be large, $y_{\eta_t}(\Lambda) = \infty\ (\geq \sqrt{4\pi})$. These represent the boundary conditions imposed on each coupling constant following the BHL approach. In the result, $\kappa_4(\mu)$ and $\kappa_5(\mu)$ are so small\footnote{Such terms arise from loop effects and tend to be small. For simplicity, we will neglect such couplings with $H$ in this section, as well as analogous couplings in the next section.}. So, $\kappa_i(\mu)$ and $\lambda_{\eta_t}(\mu)$ are llarger than $0$. Strictly speaking, the BHL compositeness condition requires
$y_{\eta_t}(\Lambda)\!\to\!\infty$. For numerical work we replace this
limit by a large but finite boundary value and set
$y_{\eta_t}(\Lambda)=\sqrt{4\pi}$. This choice (i) saturates the conventional
perturbativity/partial-wave-unitarity estimate for Yukawa couplings, and (ii) is sufficiently large that the RG flow rapidly converges
onto the same infrared compositeness trajectory as the
$y_{\eta_t}(\Lambda)\!\to\!\infty$ limit. Taking an even larger boundary
value leaves low-energy predictions unchanged within numerical accuracy.

Moreover, $\eta_t$ and $\Psi$ are constructed as doublets that, in analogy to the Higgs field, do not acquire VEV, thereby preserving gauge invariance. Also, for convenience in writing a gauge-invariant EFT, we extend $\eta_t$ by introducing a neutral scalar component and a charged component. In this work, these additional components contribute effectively only at the loop-level, and we therefore consider their effects only to that extent in what follows. Accordingly, we do not address whether the scalar or charged $\eta_t$ components could be experimentally observable.

We examine the MPP conditions at the MPP scale near $\mu\simeq \mu_c$ to determine the parameter $y_\Psi(\mu)$ as follows:
\begin{eqnarray}\label{conditions}
\left.V_{\mathrm{eff}}\right|_{\mu=\mu_c}&=&0,\\
\label{conditions2}
\left.\frac{dV_{\mathrm{eff}}}{d\mu}\right|_{\mu=\mu_c}&=&0.
\end{eqnarray}
These BHL boundary conditions and the MPP conditions are imposed as constraints on the effective potential. In our scenario, the one-loop effective potential of the SM in Landau gauge using $\mathrm{\overline{MS}}$ scheme reads\cite{higgs3}:
\begin{eqnarray}
V_{\mathrm{eff}}\left(h(\mu),\mu\right) &=& \frac{\lambda(\mu)}{4}h^4(\mu)\nonumber\\
&+&\frac{1}{64\pi^2}\left(3\lambda(\mu) h^2(\mu)\right)^2\left(\ln\frac{3\lambda(\mu) h^2(\mu)}{\mu^2}-\frac{3}{2}\right)\nonumber\\
&+&\frac{3}{64\pi^2}\left(\lambda(\mu) h^2(\mu)\right)^2\left(\ln\frac{\lambda(\mu) h^2(\mu)}{\mu^2}-\frac{3}{2}\right)\nonumber\\
&+&\frac{3\times2}{64\pi^2}\left(\frac{g_2(\mu)h(\mu)}{2}\right)^4\left(\ln\frac{\left(g_2(\mu)h(\mu)\right)^2}{4\mu^2}-\frac{5}{6}\right)\nonumber\\
&+&\frac{3}{64\pi^2}\left(\frac{\sqrt{g_2^2(\mu)+g_Y^2(\mu)}}2h(\mu)\right)^4\left(\ln\frac{\left(g_2^2(\mu)+g_Y^2(\mu)\right)h^2(\mu)}{4\mu^2}-\frac{5}{6}\right)\nonumber\\
&-&\frac{4\times3}{64\pi^2}\left(\frac{y_t(\mu)h(\mu)}{\sqrt{2}}\right)^4\left(\ln\frac{\left(y_t(\mu)h(\mu)\right)^2}{2\mu^2}-\frac{3}{2}\right)\nonumber\\
&+&\frac{4}{64\pi^2}\left(\frac{\kappa_1(\mu) h^2(\mu)}2\right)^2\left(\ln\frac{\kappa_1(\mu) h^2(\mu)}{2\mu^2}-\frac{3}{2}\right)\nonumber\\
&+&\frac{4}{64\pi^2}\left(\frac{\kappa_2(\mu) h^2(\mu)}2\right)^2\left(\ln\frac{\kappa_2(\mu) h^2(\mu)}{2\mu^2}-\frac{3}{2}\right)
\end{eqnarray}
where $h(\mu),\lambda(\mu),y_t(\mu), g_Y(\mu), g_2(\mu)$ represent the Higgs field and SM coupling constants with each depending on $\mu$ which is the renormalization scale. We focus on examining the behavior near the MPP scale $\mu_c$. However, the mass term in the Higgs field is the Electroweak scale, and is sufficiently small.
Therefore, it is neglected in this effective potential. If we require VEV of $\Psi$ to vanish, we must ensure that $\lambda_\Psi(M_t)\ge 0$ and $\kappa_1(M_t)\ge 0$. Furthermore, imposing the MPP condition along the $\Psi$ direction (i.e., the stationarity condition with respect to $\Psi$) implies not only $\lambda_\Psi(\mu)\ge 0$ and $\kappa_1(\mu)\ge 0$, but also that these couplings are sufficiently larger than the corresponding loop-induced corrections. Therefore, in this setup the MPP constraints effectively do not drive the vacuum away along the $\Psi$ axis, and it is adequate, for the purpose of the MPP analysis, to focus on the $h$ direction in field space.

We search for the parameter $ y_\Psi(\mu)$ that satisfies MPP conditions. For this reason, we examine the one-loop RGEs [\ref{firstappendix}]. Also, we simply put $h=\mu$ since the effect on the effective potential is negligibly small and $h(\mu)$ is the renormalized running field.\footnote{We do not distinguish between the bare field and the renormalized running field as its wave function renormalization is so small.}. 
We use \cite{cern,PDG}. Using the top quark mass $M_t = 172.69 \pm 0.30\ \mathrm{GeV}$ and the strong gauge coupling constant $\alpha_s(M_Z) = 0.1179 \pm 0.0010$\cite{PDG}, we numerically calculate the parameter that approximately satisfies the MPP conditions. In this case, $y_\Psi(\mu)$ indirectly affects the effective potential through the RGEs. The MPP scale is found to be $\mu_c \lesssim 10^{12.3}\ \mathrm{GeV}$. To suppress potential corrections from additional new particles, we set the upper limit to $\mu_c = 10^{12.3}\,\mathrm{GeV}$. In short, the MPP shift is minimized and unknown heavy states effects are diluted under standard EFT matching. Also, since the threshold effects persist up to several tens of GeV above the toponium mass\cite{bound}, we conservatively assume that the short-distance effects near the threshold extend over a similar energy range, and take their upper limit to be $\Lambda = 370$–$400,\mathrm{GeV}$. Moreover, since the lower bound lies around $345~\mathrm{GeV}$\cite{bound}, we assume the four-fermion interaction emerges near this scale. Accordingly, the terms involving $\eta_t$ are run between $345~\mathrm{GeV}$ and the cutoff $\Lambda$. For $\Lambda>\mu$, such a four-fermion interaction is absent, since this coupling is intended only to describe the near-threshold region. Consequently, it does not conflict with the relevant experimental constraints. The results are as follows:
\begin{eqnarray}\label{mpp}\label{11}
|y_\Psi(M_t)|\le0.3,\ \\ 0.24\le\kappa_1(M_t) \le 0.26 \quad \text{for} \ \ \ \Lambda&=&370\,\mathrm{GeV},\ \\
0.27\le\kappa_1(M_t) \le 0.29 \quad \text{for} \ \ \ \Lambda&=&400\,\mathrm{GeV}
\end{eqnarray}
where we impose the condition that $\lambda_\Psi(\mu),\ \kappa_1(\mu) \geq 0$ and $\lambda(M_t)\sim\lambda_{\Psi}(M_t)$. The effect of varying $\Lambda$ was confined to a slight change in $\kappa_1(M_t)$ and was otherwise negligible. As shown in \cite{bound_ef}, the threshold effects of toponium cannot be fully reproduced within a purely elementary-field framework. However, the change in the allowed ranges of the couplings induced by varying the cutoff above is negligible, and the discrepancy currently under consideration is of comparable size or smaller; therefore, it can be ignored for the purposes of the present discussion. We also briefly discuss the behavior when varying $\lambda_\Psi(M_t)$. As $\lambda_\Psi(M_t)$ increases, both $\kappa_1(M_t)$ and $|y_\Psi(M_t)|$ decrease slightly.
Conversely, even when $\lambda_\Psi(M_t)$ is taken smaller than $\lambda(M_t)$, $\kappa_1(M_t)$ and $|y_\Psi(M_t)|$ do not become significantly larger.
In addition, $\mu_c$ decreases as $\lambda_\Psi(M_t)$ increases, and increases as $\lambda_\Psi(M_t)$ decreases. 

\begin{figure}[H]
 \begin{center}
 \includegraphics[width=100mm]{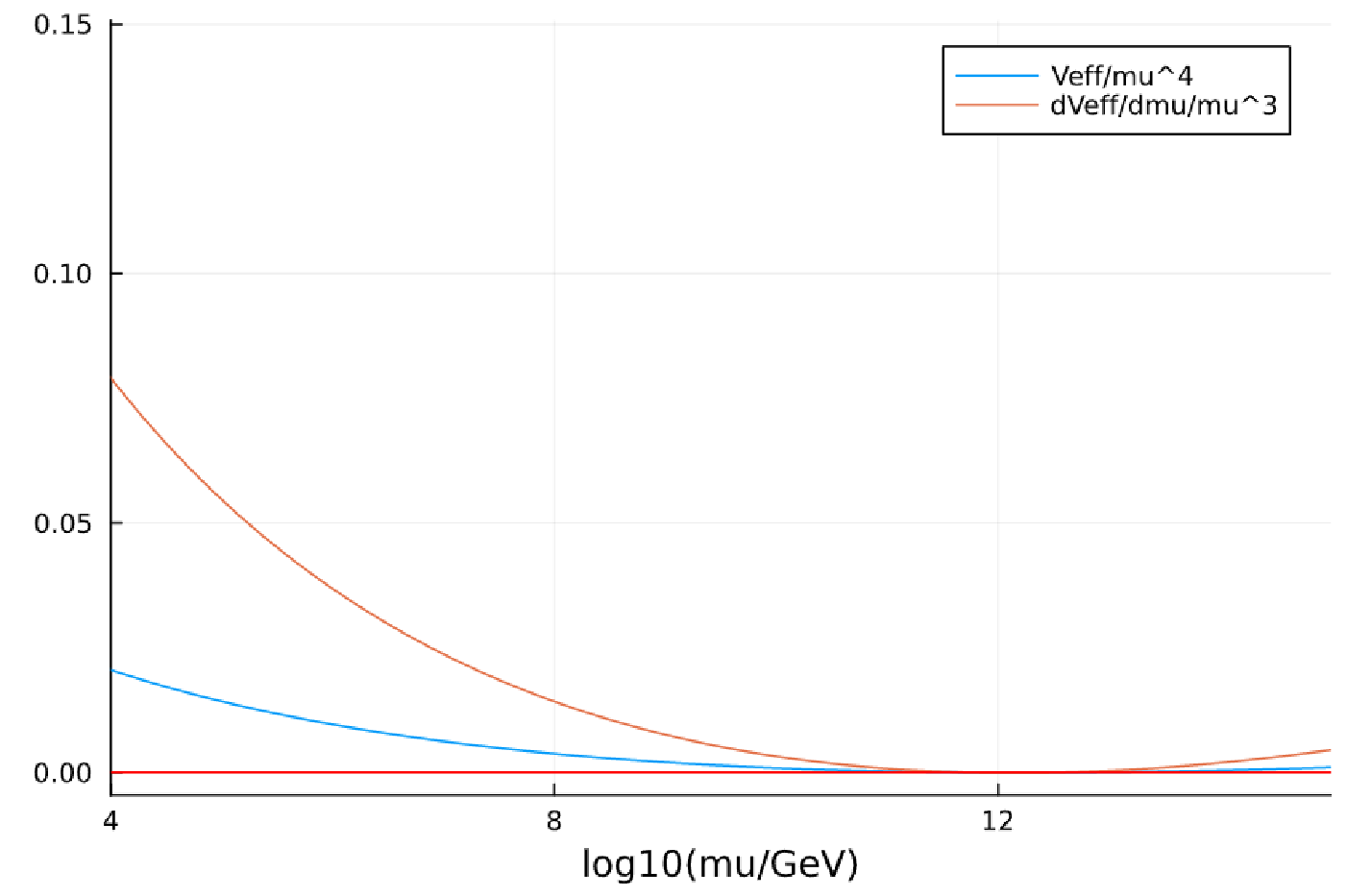}
 \caption{The $x$-axis and $y$-axis show $\log_{10}(\frac{\mu}{\mathrm{GeV}})$ and the value of each line. The blue line and orange line are $\frac{V_{\mathrm{eff}}}{\mu^4}$ and $\frac{1}{\mu^3}\frac{dV_{\mathrm{eff}}}{d\mu}$ in the case of $M_t=172.69\ \mathrm{GeV},\ \alpha_s(M_Z)=0.1189, \Lambda = 400\ \mathrm{GeV},\ y_\Psi(M_t) = 0.3,$ and $\kappa_1(M_t) =0.29$. The red line shows zero on the $y$-axis. It can be seen that the MPP condition is satisfied around $\mu_c=10^{12.3}$ GeV.}\label{fig1}
 \end{center}
\end{figure}

\begin{figure}[H]
 \begin{center}
 \includegraphics[width=100mm]{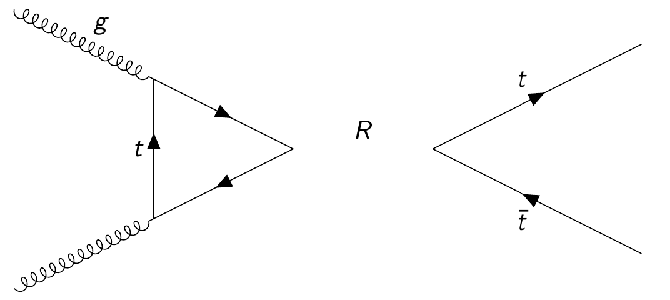}
 \caption{The production and decay process involving the mixed mass eigenstates is $gg\rightarrow R \rightarrow t\bar{t}$. $R= \eta_t^\prime,\ \Psi^\prime$}\label{fig2}
 \end{center}
\end{figure}

The behavior of the effective potential is illustrated in Figure \ref{fig1}, where it can be clearly observed that the MPP condition is satisfied in the vicinity of $\mu_c = 10^{12.3}$ GeV. Here, $\lambda(M_t)\sim\lambda_{\Psi}(M_t)$.
And the mixing angle $\theta$ between $\eta_t$ and $\Psi$ (both being pseudo-scalar components) is likewise essentially determined. Note that we are not dealing with stable particles here, so the mixing angle can be written as a complex number, $\theta=\theta_R+i\theta_C$. As will be discussed later, since we assume the decay widths are sufficiently small, we may approximate $|\theta|\sim|\theta_R|$; hence the allowed range is $|\theta|\le45^\circ$, and it is determined as follows:
\begin{eqnarray}
\begin{pmatrix}
\eta_t^\prime \\
\Psi^\prime
\end{pmatrix}
=
\begin{pmatrix}
\cos\theta & -\sin\theta \\
\sin\theta & \cos\theta
\end{pmatrix}
\begin{pmatrix}
\eta_t \\
\Psi
\end{pmatrix}.
\end{eqnarray}
In this case, the mass matrix $\mathcal{M}$ takes the following form:
\begin{eqnarray}
\mathcal{M}
=
\begin{pmatrix}
m_{\eta_t}^2-im_{\eta_t}\Gamma_{\eta_t}(\equiv A) & \delta \\
\delta & m_{\Psi}^2-im_{\Psi}\Gamma_{\Psi} (\equiv D)
\end{pmatrix}
\end{eqnarray}
where $\Gamma_i$ is the total decay widths of each state and $\delta$ is an off-diagonal elements of the mass matrix. So, $\theta$ can be expressed as follows:
\begin{eqnarray}
\tan 2\theta = \frac{2\delta}{A-D}.
\end{eqnarray}
When mass mixing with a new field $\Psi$ is introduced, mass eigenstates emerge, and the off-diagonal components of the mass matrix constrain the mixing angle. At this stage, mixing can allow the toponium-like eigenstate to acquire a lifetime different from its formation timescale. Figure \ref{fig2} illustrates representative production and decay processes for the two mixed eigenstates, with $R$ denoting $\eta_t^\prime$ or $\Psi^\prime$. In addition to $t\bar{t}$ production, $gg$ and $\gamma\gamma$ production channels are also possible for $R$.
We further assume that the total decay widths of each state, as well as their differences, are sufficiently small compared to their respective masses, such that about $\frac{\Gamma_\Psi}{m_\Psi} = 2\%$. In addition, the mass of toponium is taken to be approximately $345 \ \mathrm{GeV}$, and it is assumed that, even after mixing with $\Psi$, one of the mass eigenvalues remains close to $345 \ \mathrm{GeV}$ with negligible shift.

We focus on the scenario in which a new scalar and pseudo-scalar share the same Yukawa coupling constant, both with a mass of approximately $365 \  \mathrm{GeV}$ and a relative decay width of approximately $2\%$. Taking into account the CMS results\cite{toponium}, from the experimentally determined exclusion limits and the deviation of numerically calculated observations-both extending beyond the $2\sigma$ range-the allowed region of the Yukawa coupling of the mixed state, $y_{\Psi^\prime}(M_t)$, is obtained as follows:
\begin{eqnarray}\label{222}
0.4\le |y_{\Psi^\prime}(M_t)| \le 0.5.
\end{eqnarray}

Correspondingly, the mixing angle must satisfy
\begin{eqnarray}
|\theta| \le 13^\circ.
\end{eqnarray}
where assuming the imaginary part is sufficiently small, we determine the mixing angle from the bounds given by Eqs. (\ref{11}) and (\ref{222}). It should be noted that $y_{\eta_t}$ does not exist at $\mu = M_t$. However, since $y_{\Psi}(M_t)$ and $y_{\Psi}(\mu = 345~\mathrm{GeV})$ differ only negligibly for any value of $y_{\Psi}(M_t)$, they can be approximated to be nearly identical.

In the setup adopted here, there remain three assumptions that have not been explained explicitly: (i) the assumed decay widths of the new states, (ii) the assumption that the new scalars share identical Yukawa couplings, and (iii) the requirement that one of the mass eigenstates lies near 
$345$ GeV. The first assumption is introduced mainly as a simplification to realize $|\theta|\sim|\theta_R|$. The second follows from the fact that the new fields form the $SU(2)_L$ doublet (and is therefore more a natural consequence than an assumption). The third is imposed so that the scenario remains consistent with current experimental observations.
\section{In case of the 2HDM to explain the excess in LHC}\label{sec4}
In this Section, we consider the model in which the doublet $\eta_t$ discussed above is added to the 2HDM.
The additional tree scalar potential is
\begin{eqnarray}
V(H_1,H_2,\eta_t) &=& M_{H_1}^2(H_1^\dagger H_1)+M_{H_2}^2(H_2^\dagger H_2)+M_{H_1H_2}^2(H_1^\dagger H_2+\mathrm{h.c.}))\nonumber\\
&+&M_{\eta_t}^2(\eta_t^\dagger\eta_t)+M_{H_1\eta_t}^2(H_1^\dagger \eta_t+\mathrm{h.c.})+M_{H_2\eta_t}^2(H_2^\dagger \eta_t+\mathrm{h.c.})\nonumber\\
&+&\frac{\lambda_1}2(H_1^\dagger H_1)^2+\frac{\lambda_{2}}2(H_2^\dagger H_2)^2+\frac{\lambda_{\eta_t}}2(\eta_t^\dagger \eta_t)^2\nonumber\\
&+&\lambda_{3}(H_1^\dagger H_1)(H_2^\dagger H_2)+\kappa_{6}(H_1^\dagger H_1)(\eta_t^\dagger \eta_t)\nonumber\\
&+&\kappa_{7}(H_2^\dagger H_2)(\eta_t^\dagger \eta_t)+\lambda_{4}(H_1^\dagger H_2)(H_2^\dagger H_1)\nonumber\\
&+&\frac{\lambda_{5}}2((H_1^\dagger H_2)^2+\mathrm{h.c.})
\end{eqnarray}
where we additionally introduce $\kappa_6$ and $\kappa_7$, and set the BHL boundary conditions as $\kappa_6(\Lambda) = \kappa_7(\Lambda) = \lambda_{\eta_t}(\Lambda) = 0$ and in order to avoid mixing with the SM Higgs, we set $M^2_{H_1\eta_t} \sim -M^2_{H_2\eta_t}\tan\beta$. This is in order to evade the constraints from the scalar mixing angle with the SM Higgs at the LHC and from the Higgs mass. In other words, it mixes with other fields in the non-SM sector.
 
Assuming, for simplicity, all coupling constants are real. The MPP conditions can be satisfied at any scale $\mu_c$ (below the Planck scale) and $\Lambda = 370$--$400$ GeV and the lower bound is around $345~\mathrm{GeV}$ as follows:
\begin{eqnarray}\label{mpp2hdm}
|\lambda_4(M_t)| \le 0.5,\quad |\lambda_5(M_t)| \le 0.5\quad\ \text{for } \ 370\leq\Lambda\leq400\,\mathrm{GeV}
\end{eqnarray}
where we use the one-loop effective potential [\ref{secondappendix}] and the one-loop \\ RGEs [\ref{thirdappendix}]. And it satisfies 
\begin{eqnarray}\label{mpp2hdm2}
\lambda_1(M_t)+\lambda_2(M_t)\sim2\lambda_3(M_t).
\end{eqnarray}
When this condition is approximately satisfied, the new scalar field and the pseudo-scalar field acquire nearly degenerate masses.

In addition, the vacuum stability conditions are given by
\begin{eqnarray}
\lambda_1> 0,\quad \lambda_2 > 0,\quad \lambda_3> -\sqrt{\lambda_1\lambda_2},\quad \lambda_3 + \lambda_4 - |\lambda_5| > -\sqrt{\lambda_1\lambda_2}.
\end{eqnarray}
These vacuum stability conditions are also satisfied by Eq.~(\ref{mpp2hdm}). Moreover, the Higgs alignment condition is satisfied at $\mu=M_t$.

It should be noted that the so-called oblique parameters ($S$, $T$, and $U$) must be taken into account between the 2HDM and $\eta_t$\cite{2hdm}. After the basis transformation in the 2HDM, mixing arises between mass component (the new pseudo-scalar) and the corresponding component (pseudo-scalar) of $\eta_t$. However, if we assume that the mass of toponium remains approximately $345$ GeV even after the mixing, the constraints from the oblique parameters can be neglected, since in such a case the mass of component (the new pseudo-scalar) on the 2HDM side also change only negligibly. Therefore, it is sufficient to consider only the deviations of the oblique parameters originating from the 2HDM.

The current experimental constraints on the various types of the 2HDM, namely Type II and Y, are imposed on the charged Higgs boson mass $M_{H^\pm}$ and the mixing angle in the Higgs basis, $\tan\beta$\cite{DM2}.  
For Type II and Y, the bound\cite{Obs4,Obs5} is
\begin{eqnarray}
M_{H^\pm} \gtrsim 800\ \mathrm{GeV}, \quad \text{for any } \tan\beta.
\end{eqnarray}

In this context, the masses of the pseudo-scalar $A$ and the charged scalar $H^\pm$ arising from the 2HDM are given by the following relations, using $v_{\text{EW}} = 246\,\mathrm{GeV}$:
\begin{eqnarray}
M_{A}^2 &=& \frac{M_{H_1H_2}^2}{\sin\beta\cos\beta} - \lambda_5 v_{\text{EW}}^2, \\
M_{H^\pm}^2 &=& \frac{M_{H_1H_2}^2}{\sin\beta\cos\beta} - \frac{\lambda_4 + \lambda_5}{2} v_{\text{EW}}^2. 
\end{eqnarray}

From Eq (\ref{mpp2hdm}), it follows that in Type II and Type Y:
\begin{eqnarray}
M_{A} \gtrsim 800\,\mathrm{GeV}.
\end{eqnarray}
The oblique parameters can be kept consistent with observations and within the experimental bounds under such conditions. In fact, when Eqs.~(\ref{mpp2hdm}) and (\ref{mpp2hdm2}) are satisfied, the mass differences among each component are at most several tens of GeV, and within this range, they do not affect the constraints from the oblique parameters\cite{2hdm}.

In this setup, the mixing angle must satisfy
\begin{eqnarray}
|\theta| \le 1^\circ.
\end{eqnarray}
Here, we work with the mass matrix and assume that the off-diagonal $\delta m$ entries are not large (we take $(|\delta m^2|\lesssim (100~\mathrm{GeV})^2$)). Including the values of the diagonal elements of the mass matrix as well ($\gtrsim 800\,\mathrm{GeV}$ and $\sim 345\,\mathrm{GeV}$), we determined the mixing angle.
This assumption is based on the fact that these entries originate from the top-quark Yukawa coupling and are of the same order in magnitude as it. In other words, short-distance top-loop effects make a sizable radiative contribution to the off-diagonal mass terms and keep their size moderate. In conjunction with the condition $M^2_{H_1\eta_t}\sim -M^2_{H_2\eta_t}\tan\beta$, the off-diagonal entries are thereby fixed. That is, their magnitude is effectively controlled to be of the same order as the loop effects.\footnote{It is, of course, possible to introduce large tree-level off-diagonal mass terms by hand so that $M^2_{H_1\eta_t}\sim -M^2_{H_2\eta_t}\tan\beta$ holds. However, enforcing this relation entails fine-tuning and is therefore quite unnatural. If one allows for a large tree-level off-diagonal mass term and performs an appropriate fine-tuning, the mixing angle may exceed $|\theta|\le1^\circ$. For this reason, it is important to discuss the role of fine-tuning in the present framework.}
However, such a small mixing angle is highly unnatural.
Indeed, $|\theta|\le 1^\circ$ corresponds to (near) the decoupling limit, and in the absence of a symmetry that protects $\theta \to 0$, this smallness is (technically) unnatural\footnote{In the minimal scenario, the denominator $A-D$ in the relation for $\tan 2\theta$ becomes small, so this kind of unnaturalness does not occur.}.

Furthermore, $M_{A} \gtrsim 800\,\mathrm{GeV}$ is disfavored by the CMS results.


\section{Summary and Discussion}\label{sec5}
Motivated by the excess of $t\bar t$ events near threshold reported by the LHC experiments, we studied a framework in which a bound state of top quarks (toponium, $\eta_t$) mixes with an additional elementary field $\Psi$. Our analysis was carried out in an effective description that incorporates short-distance effects near the threshold via the Bardeen-Hill-Lindner (BHL) prescription, supplemented with the Multicritical Point Principle (MPP) as a high-scale constraint on couplings.

Within the BHL framework we impose a compositeness boundary condition on the wavefunction renormalization of $\eta_t$ at a cutoff $\Lambda$, and use MPP to restrict the parameter space of the low-energy EFT. We considered (i) a minimal scenario where $\Psi$ mixes with $\eta_t$ but has negligible couplings to other fermions, and (ii) an embedding into Two-Higgs-Doublet Models (2HDM, Types~II and Y) augmented by $\eta_t$.

In the minimal scenario the mixing angle between $\Psi$ and $\eta_t$ is bounded by
$|\theta|\le 13^\circ$, compatible with the $t\bar t$ excess and the MPP-implied parameter range. When recast into the 2HDM (Types~II and Y), the combined MPP and vacuum-stability requirements, together with charged-Higgs searches, enforce $M_A \gtrsim 800~\mathrm{GeV}$ and drive the mixing to the decoupling regime, yielding a much tighter limit $|\theta|\le 1^\circ$.

The extremely small mixing required in the 2HDM embedding is not particularly desirable, being (technically) unnatural. Short-distance dynamics associated with toponium formation radiatively generate the mixing through top-loop corrections while preventing it from becoming excessively large. Consequently, $|\theta|\le 1$ follows; however, in the absence of a protective symmetry, maintaining this smallness is unnatural.

Throughout, we assumed (i) a toponium mass close to $345~\mathrm{GeV}$ even after mixing and (ii) relatively small total widths ($\Gamma/M = 2\%$). Under these assumptions, mass splittings among new scalar, pseudo-scalar, and charged states remain modest, so oblique-parameter constraints are dominated by the 2HDM sector and are not further tightened by $\eta_t$–mixing effects. These assumptions should be borne in mind when interpreting the allowed parameter space.

If the observed $t\bar t$ excess originates from $\eta_t$-$\Psi$ mixing, the minimal scenario is both less constrained and more predictive than a generic 2HDM embedding, owing to its smaller coupling set and effectively negligible interactions with non-top fermions. Within the presently allowed ranges of couplings and masses, however, the resulting bounds are not substantially different from those obtained in 2HDM Types I and X for $\tan\beta \ge 1.0$. Compared to 2HDM Types I and X, the minimal scenario is an attractive target for near-term probes owing to its lower sensitivity to constraints from flavor physics, charged Higgs searches, and lepton-related bounds, as well as its smaller number of degrees of freedom. Conversely, whether the scenario conflicts with constraints from flavor physics, charged Higgs searches, or lepton-related bounds provides a powerful way to test and distinguish the minimal scenario from the 2HDM of Types I and X. A decisive assessment will nevertheless require: (i) refined collider studies of threshold line shapes and associated channels ($gg$, $\gamma\gamma$) for the two mixed eigenstates \cite{fu1,fu2}, (ii) incorporation of higher-loop corrections in the RG and effective-potential analyses \cite{fu3,fu4}, and (iii) a systematic survey of electroweak precision and flavor constraints beyond the leading approximations employed here \cite{fu5}.

Taken together, our results favor the minimal mixing scenario as a phenomenologically viable and predictive explanation of the threshold $t\bar t$ excess, while the 2HDM embedding tends to suppress observable mixing effects. Future collider measurements and higher-order theoretical inputs will be crucial to sharpen these conclusions and to further test the link between the $t\bar t$ excess and physics beyond the Standard Model.

\section*{Acknowledgments}
We thank Noriaki Aibara, Gi-Chol CHO, So Katagiri, Shiro Komata, and Akio Sugamoto for many helpful comments. Especially, I would like to take this opportunity to extend my deepest appreciation to Gi-Chol CHO and Akio Sugamoto for their generous advice and the fruitful discussions we shared.

\appendix
\renewcommand{\thesection}{Appendix \Alph{section}}
\section{One-loop Renormalization Group Equations including a new field and toponium}\label{firstappendix}
The one-loop RGEs are
\begin{eqnarray}
\frac{dg_Y}{dt}&=&\frac{g_Y^3}{16\pi^2}(7+\frac{1}{6}\delta),\\ \frac{dg_2}{dt}&=&\frac{g_2^3}{16\pi^2}(-3+\frac{1}{6}\delta),\\ \frac{dg_3}{dt}&=&\frac{g_3^3}{16\pi^2}(-7), \\ 
\frac{dy_t}{dt}&=&\frac{y_t}{16\pi^2}\Big(\frac{9}2y_t^2+\frac{3}2y_\Psi^2+\frac{3}2y_{\eta_t}^2-\frac{17}{12}g_Y^2-\frac{9}4g_2^2-8g_3^2\Big),\\
\frac{dy_\Psi}{dt}&=&\frac{y_\Psi}{16\pi^2}\Big(\frac{9}2y_\Psi^2+\frac{3}2y_t^2+\frac{3}2y_{\eta_t}^2-\frac{17}{12}g_Y^2-\frac{9}4g_2^2-8g_3^2\Big),\\
\frac{dy_{\eta_t}}{dt}&=&\frac{y_{\eta_t}}{16\pi^2}\Big(\frac{9}2y_{\eta_t}^2+\frac{3}2y_t^2+\frac{3}2y_{\Psi}^2-\frac{17}{12}g_Y^2-\frac{9}4g_2^2-8g_3^2\Big),\\
\frac{d\lambda}{dt} &=& \frac{1}{16\pi^2}\Big(\lambda\left(24\lambda-3g_Y^2-9g_2^2+12y_t^2\right)\nonumber \\
&+&2\kappa_{1}^2+2\kappa_{2}^2+\frac{3}{8}g_Y^4+\frac{3}{4}g_Y^2g_2^2+\frac{9}{8}g^4_2-6y_t^4\Big),\\
\frac{d\kappa_{1}}{dt} &=& \frac{1}{16\pi^2}\Big(\kappa_{1}\left(4\kappa_{1}+12\lambda+6\lambda_\Psi-3g_Y^2-9g_2^2+6y_t^2+6y_\Psi^2\right)\nonumber\ \\
&+&4\kappa_2\kappa_3+\frac{3}{4}g_Y^4-\frac{3}{2}g_Y^2g_2^2+\frac{9}{4}g^4_2-6y_t^2y_\Psi^2\Big),\ \\
\frac{d\kappa_{2}}{dt} &=& \frac{1}{16\pi^2}\Big(\kappa_{2}\left(4\kappa_{2}+12\lambda+6\lambda_{\eta_t}-3g_Y^2-9g_2^2+6y_t^2+6y_{\eta_t}^2\right)\nonumber\ \\
&+&4\kappa_1\kappa_3+\frac{3}{4}g_Y^4-\frac{3}{2}g_Y^2g_2^2+\frac{9}{4}g^4_2-6y_t^2y_{\eta_t}^2\Big),\ \\
\frac{d\kappa_{3}}{dt} &=& \frac{1}{16\pi^2}\Big(\kappa_{3}\left(4\kappa_{3}-3g_Y^2-9g_2^2+6y_\Psi^2+6y_{\eta_t}^2\right)+(\lambda_\Psi+\lambda_{\eta_t})(6\kappa_3+2\kappa_4)\nonumber\ \\
&+&4\kappa_1\kappa_2+2\kappa_4^2+2\kappa_5^2+\frac{3}{4}g_Y^4-\frac{3}{2}g_Y^2g_2^2+\frac{9}{4}g^4_2-6y_\Psi^2y_{\eta_t}^2\Big),\ \\
\frac{d\kappa_{4}}{dt} &=& \frac{1}{16\pi^2}\Big(\kappa_{4}\left(4\kappa_{4}+8\kappa_3-3g_Y^2-9g_2^2+6y_t^2+6y_\Psi^2\right)\nonumber\ \\
&+&2(\lambda_\Psi+\lambda_{\eta_t})\kappa_4+8\kappa_5^2+3g_Y^2g_2^2-6y_\Psi^2y_{\eta_t}^2\Big),\ \\
\frac{d\kappa_{5}}{dt} &=& \frac{1}{16\pi^2}\Big(\kappa_{5}\left(12\kappa_{4}+8\kappa_3-3g_Y^2-9g_2^2+6y_t^2+6y_\Psi^2\right)\nonumber\ \\
&+&2(\lambda_\Psi+\lambda_{\eta_t})\kappa_5-12y_\Psi^2y_{\eta_t}^2\Big),\ \\
\frac{d\lambda_\Psi}{dt} &=& \frac{1}{16\pi^2}\Big(\lambda_\Psi\left(12\lambda_\Psi-3g_Y^2-9g_2^2+12y_\Psi^2\right)\nonumber\ \\
&+&4\kappa_{1}^2+4\kappa_{3}^2+4\kappa_{3}\kappa_4+2\kappa_4^2+2\kappa_5^2+\frac{3}{4}g_Y^4+\frac{3}{2}g_Y^2g_2^2+\frac{9}{4}g^4_2-12y_\Psi^4\Big),\ \\
\frac{d\lambda_{\eta_t}}{dt} &=& \frac{1}{16\pi^2}\Big(\lambda_{\eta_t}\left(12\lambda_{\eta_t}-3g_Y^2-9g_2^2+12y_{\eta_t}^2\right)\nonumber\ \\
&+&4\kappa_{2}^2+4\kappa_{3}^2+4\kappa_{3}\kappa_4+2\kappa_4^2+2\kappa_5^2+\frac{3}{4}g_Y^4+\frac{3}{2}g_Y^2g_2^2+\frac{9}{4}g^4_2-12y_{\eta_t}^4\Big)
\end{eqnarray}
where $t=\ln\mu$. $\mu$ is the renormalization scale. $\delta=1$ when $345\ \mathrm{GeV}\le\mu\le\Lambda$ and $\delta=0$ otherwise. $\delta$ is set to $1$ in the region where the one-loop contribution of $\eta_t$ is present.

\section{Two-Higgs-Doublet Models One-loop Effective Potential including Toponium}\label{secondappendix}
The one-loop effective potential is
\begin{eqnarray}
V(h_1(\mu),h_2(\mu),\mu)&=&\frac{\lambda_1(\mu)}8h_1^4(\mu)+\frac{\lambda_{2}(\mu)}8h_2^4(\mu)+(\frac{\lambda_3(\mu)}4+\frac{\lambda_4(\mu)}4+\frac{\lambda_5(\mu)}8)h^2_1(\mu)h_2^2(\mu)\nonumber\\
&+&\frac{1}{64\pi^2}\mathrm{tr}(M_i^2(\mu))^2\left(\ln\frac{M_i^2(\mu)}{\mu^2}-\frac{3}{2}\right)\nonumber\\
&+&\frac{3\times2}{64\pi^2}\left(\frac{g_2(\mu)h(\mu)}{2}\right)^4\left(\ln\frac{\left(g_2(\mu)h(\mu)\right)^2}{4\mu^2}-\frac{5}{6}\right)\nonumber\\
&+&\frac{3}{64\pi^2}\left(\frac{\sqrt{g_2^2(\mu)+g_Y^2(\mu)}}2h(\mu)\right)^4\left(\ln\frac{\left(g_2^2(\mu)+g_Y^2(\mu)\right)h^2(\mu)}{4\mu^2}-\frac{5}{6}\right)\nonumber\\
&-&\frac{4\times3}{64\pi^2}\left(\frac{y_t(\mu)h_2(\mu)}{\sqrt{2}}\right)^4\left(\ln\frac{\left(y_t(\mu)h_2(\mu)\right)^2}{2\mu^2}-\frac{3}{2}\right)\nonumber \\
&+&\frac{4}{64\pi^2}\left(\frac{\kappa_{6}^2(\mu)}4h_1(\mu)^4\right)\left(\ln\frac{\kappa_{6}(\mu)h_1^2(\mu)}{2\mu^2}-\frac{3}{2}\right)\nonumber\\
&+&\frac{4}{64\pi^2}\left(\frac{\kappa_{7}^2(\mu)}4h_2(\mu)^4\right)\left(\ln\frac{\kappa_{7}(\mu)h_2^2(\mu)}{2\mu^2}-\frac{3}{2}\right)
\end{eqnarray}
where $h_1^2(\mu)+h_2^2(\mu) \equiv h^2(\mu)$ and\ \\ $i=S,P,C, \lambda_{345}(\mu)=\lambda_3(\mu)+\lambda_4(\mu)+\lambda_5(\mu)$,\ \\$M_S^2=\begin{pmatrix}
\lambda_1(\mu)h_1^2(\mu) & \lambda_{345}(\mu)h_1(\mu)h_2(\mu)  \\
\lambda_{345}(\mu)h_1(\mu)h_2(\mu) & \lambda_2(\mu)h_2^2(\mu)
\end{pmatrix}$, \ \\$M_P^2=\begin{pmatrix}
-\lambda_5(\mu)h_2^2(\mu) & \lambda_5(\mu)h_1(\mu)h_2(\mu) \\
\lambda_5(\mu)h_1(\mu)h_2(\mu) & -\lambda_5h_1^2(\mu)
\end{pmatrix}$,\\ $M_C^2=\begin{pmatrix}
-\frac{\lambda_4(\mu)+\lambda_5(\mu)}2h_2^2(\mu) & \frac{\lambda_4(\mu)+\lambda_5(\mu)}2h_1(\mu)h_2(\mu) \\
\frac{\lambda_4(\mu)+\lambda_5(\mu)}2h_1(\mu)h_2(\mu) &-\frac{\lambda_4(\mu)+\lambda_5(\mu)}2h_1^2(\mu)
\end{pmatrix}$.
\section{Two-Higgs-Doublet Models One-loop Renormalization Group Equations including Toponium}\label{thirdappendix}
The one-loop RGEs are
\begin{eqnarray}
\frac{dg_Y}{dt}&=&\frac{g_Y^3}{16\pi^2}(7+\frac{1}{6}\delta),\\ \frac{dg_2}{dt}&=&\frac{g_2^3}{16\pi^2}(-3+\frac{1}{6}\delta),\\ \frac{dg_3}{dt}&=&\frac{g_3^3}{16\pi^2}(-7), \\ 
\frac{dy_t}{dt}&=&\frac{y_t}{16\pi^2}\Big(\frac{9}2y_t^2+\frac{3}2y_{\eta_t}^2-\frac{17}{12}g_Y^2-\frac{9}4g_2^2-8g_3^2\Big),\\
\frac{dy_{\eta_t}}{dt}&=&\frac{y_{\eta_t}}{16\pi^2}\Big(\frac{9}2y_{\eta_t}^2+\frac{3}2y_t^2-\frac{17}{12}g_Y^2-\frac{9}4g_2^2-8g_3^2\Big),\ \\
\frac{d\lambda_1}{dt} &=& \frac{1}{16\pi^2}\Big(\lambda_1\left(12\lambda_1-3g_Y^2-9g_2^2\right)\nonumber\ \\
&+&4\lambda_{3}^2+4\lambda_{3}\lambda_4+2\lambda_{4}^2+2\lambda_{5}^2+4\kappa_6^2+\frac{3}{4}g_Y^4+\frac{3}{2}g_Y^2g_2^2+\frac{9}{4}g^4_2\Big),\ \\
\frac{d\lambda_2}{dt} &=& \frac{1}{16\pi^2}\Big(\lambda_2\left(12\lambda_2-3g_Y^2-9g_2^2+12y_t^2\right)\nonumber\ \\
&+&4\lambda_{3}^2+4\lambda_{3}\lambda_4+2\lambda_{4}^2+2\lambda_{5}^2+4\kappa_7^2+\frac{3}{4}g_Y^4+\frac{3}{2}g_Y^2g_2^2+\frac{9}{4}g^4_2-12y_t^4\Big),\ \\
\frac{d\lambda_3}{dt} &=& \frac{1}{16\pi^2}\Big(\lambda_3\left(3g_Y^2-9g_2^2+6y_t^4\right)+(\lambda_1+\lambda_2)(6\lambda_3+2\lambda_4)\nonumber\ \\ &+&4\lambda_{3}^2+2\lambda_{4}^2+2\lambda_{5}^2+4\kappa_6\kappa_7+\frac{3}{4}g_Y^4-\frac{3}{2}g_Y^2g_2^2+\frac{9}{4}g^4_2\Big),\ \\
\frac{d\lambda_4}{dt} &=& \frac{1}{16\pi^2}\Big(\lambda_4\left(2\lambda_4+3g_Y^2-9g_2^2+6y_t^2\right)+2(\lambda_1+\lambda_2)\lambda_4\nonumber\ \\
&+&8\lambda_{3}\lambda_4+4\lambda_{4}^2+8\lambda_{5}^2+3g_Y^2g_2^2\Big),\ \\
\frac{d\lambda_5}{dt} &=& \frac{1}{16\pi^2}\Big(\lambda_5\left(3g_Y^2-9g_2^2+6y_t^4\right)+2(\lambda_1+\lambda_2+4\lambda_3+6\lambda_4)\lambda_5\Big),\ \\
\frac{d\lambda_{\eta_t}}{dt} &=& \frac{1}{16\pi^2}\Big(\lambda_{\eta_t}\left(12\lambda_{\eta_t}-3g_Y^2-9g_2^2+12y_{\eta_t}^2\right)\nonumber\ \\
&+&4\kappa_{6}^2+4\kappa_{7}^2+\frac{3}{4}g_Y^4+\frac{3}{2}g_Y^2g_2^2+\frac{9}{4}g^4_2-12y_{\eta_t}^4\Big),\ \\
\frac{d\kappa_{6}}{dt} &=& \frac{1}{16\pi^2}\Big(\kappa_{6}\left(4\kappa_{6}+6\lambda_{\eta_t}+6\lambda_{1}-3g_Y^2-9g_2^2+6y_{\eta_t}^2\right)\nonumber\ \\
&+&4\kappa_{7}\lambda_3+\frac{3}{4}g_Y^4-\frac{3}{2}g_Y^2g_2^2+\frac{9}{4}g^4_2\Big),\ \\
\frac{d\kappa_{7}}{dt} &=& \frac{1}{16\pi^2}\Big(\kappa_{7}\left(4\kappa_{7}+6\lambda_{\eta_t}+6\lambda_{2}-3g_Y^2-9g_2^2+6y_t^2+6y_{\eta_t}^2\right)\nonumber\ \\
&+&4\kappa_{6}\lambda_3+\frac{3}{4}g_Y^4-\frac{3}{2}g_Y^2g_2^2+\frac{9}{4}g^4_2-6y_t^2y_{\eta_t}^2\Big)
\end{eqnarray}

where $t=\ln\mu$. $\mu$ is the renormalization scale. $\delta=1$ when $345\ \mathrm{GeV}\le\mu\le\Lambda$ and $\delta=0$ otherwise. $\delta$ is set to $1$ in the region where the one-loop contribution of $\eta_t$ is present.



\begin{thebibliography}{999}
\bibitem{higgs2}S. Chatrchyan et al. [CMS Collaboration], ``Observation of a new boson at a mass of 125 GeV with the CMS experiment at the LHC,'' Phys. lett. B 716, 30 (2012), arXiv:1207.7235 [hep-ex].

\bibitem{LHC2}X. C. Vidal  et al., ``Beyond the Standard Model Physics at the HL-LHC and HE-LHC,'' (2018), arXiv:1812.07831 [hep-ph].

\bibitem{LHC3}J. Alimena et al., ``Searching for long-lived particles beyond the Standard Model at the Large Hadron Collider,'' J. Phys. G: Nucl. Part. Phys. 47 090501 (2020), arXiv:1903.04497 [hep-ex].

\bibitem{LHC1}A.-M. Lyon, Review of searches for new physics at CMS, in 58th Rencontres de Moriond on QCD and High Energy Interactions, 6, (2024), arXiv:2406.02010 [hep-ex].

\bibitem{topo_1}[ATLAS Collaboration], ``Observation of quantum entanglement in top-quark pairs using the ATLAS detector,'' Nature 633 542 (2024), arXiv:2311.07288 [hep-ex].

\bibitem{topo_2}[CMS Collaboration], ``Differential cross section measurements for the production of top quark pairs and of additional jets using dilepton events from $pp$ collisions at $\sqrt{s} = 13$ TeV,'' JHEP 02 064 (2025), arXiv:2402.08486 [hep-ex].

\bibitem{topo_3}[CMS Collaboration], ``Observation of quantum entanglement in top quark pair production in proton-proton collisions at $\sqrt{s} = 13$ TeV,'' Rep. Prog. Phys. 87 117801 (2024), arXiv:2406.03976 [hep-ex].

\bibitem{topo_4}[CMS Collaboration], ``Measurements of polarization and spin correlation and observation of entanglement in top quark pairs using lepton+jets events from proton-proton collisions at $\sqrt{s} = 13$ TeV,'' Phys. Rev. D 110, 112016 (2024), arXiv:2409.11067 [hep-ex].

\bibitem{toponium}[CMS Collaboration], ``Search for heavy pseudoscalar and scalar bosons decaying to top quark pairs in proton-proton collisions at $\sqrt{s}= 13 \mathrm{TeV}$,''  CMS PAS HIG-22-013, Rep. Prog. Phys. 88 127801 (2025), arXiv:2507.05119 [hep-ex].
\url{https://cds.cern.ch/record/2911775}

\bibitem{topo_5}[CMS Collaboration], ``Observation of a pseudoscalar excess at the top quark pair production threshold,'' Rep. Prog. Phys. 88 087801 (2025), arXiv:2503.22382 [hep-ex].

\bibitem{toponium_atlas}[ATLAS Collaboration], ``Observation of a cross-section enhancement near the $t\bar{t}$ production threshold in $\sqrt{s} = 13$ TeV $pp$ collisions with the ATLAS detector,''  ATLAS-CONF-2025-008 (2025). 
\url{https://cds.cern.ch/record/2883796}


\bibitem{toponium_1}B. Fuks, K. Hagiwara, K. Ma, and Y.-J. Zheng, ``Signatures of toponium formation in LHC run 2 data,'' Phys. Rev. D 104 034023 (2021), arXiv:2102.11281 [hep-ph]. 


\bibitem{topotopo3}Y. Matsuoka, ``On the Effects of Elementary and Bound State Fields on Vacuum Stability in $t\bar{t}$ Production at the LHC,'' Phys. lett. B 872 140033 (2026), arXiv:2410.04672 [hep-ph].


\bibitem{topotopo}Y. Nambu and G. Jona-Lasinio, ``Dynamical model of elementary particles based on an analogy with superconductivity. I,'' Phys. Rev. 122, 345 (1961).

\bibitem{topotopo2}W. A. Bardeen, C. T. Hill, and M. Lindner, ``Minimal dynamical symmetry breaking of the standard model,'' Phys. Rev. D 41, 1647 (1990).


\bibitem{MPPoriginal}C. D. Froggatt and H. B. Nielsen, ``Standard model criticality prediction: Top
mass 173 $\pm$ 5-GeV and Higgs mass 135 $\pm$ 9-GeV,'' Phys. Lett. B 368, 96(1996), [hep-ph/9511371].

\bibitem{MPPoriginal2}C. D. Froggatt, H. B. Nielsen, and Y. Takanishi, ``Standard model Higgs boson
mass from borderline metastability of the vacuum,'' Phys. Rev. D 64, 113014 (2001), [hep-ph/0104161].

\bibitem{MPPoriginal3}H. B. Nielsen, ``PREdicted the Higgs Mass,'' Bled Workshops Phys. 13 no. 2, 94–126, (2012), arXiv:1212.5716 [hep-ph].

\bibitem{kk1}K. Kawana, ``Multiple Point Principle of the Standard Model with Scalar Singlet Dark Matter and Right Handed Neutrinos,'' PTEP, 023B04 (2015), arXiv:1411.2097 [hep-ph].


\bibitem{mppeg1}N. Haba, H. Ishida, N. Okada, and Y. Yamaguchi, ``Multiple-point principle with a scalar singlet extension of the Standard Model,'' PTEP, 013B03 (2017), arXiv:1608.00087 [hep-ph].



\bibitem{kawai1}Y. Hamada, H. Kawai, K. Kawana, K.-y. Oda, and K. Yagyu, ``Minimal Scenario of
Criticality for Electroweak Scale, Neutrino Masses, Dark Matter, and Inflation,'' Eur. Phys. J. C 81 962 (2021), arXiv:2102.04617 [hep-ph].

\bibitem{mppeg2} G.-C. Cho, C. Idegawa, and R. Sugihara, ``A complex singlet extension of the Standard Model and Multi-critical Point Principle,'' Phys. lett. B 839 137757 (2023), arXiv:2212.13029 [hep-ph].

\bibitem{me1}Y. Matsuoka, ``Prediction of the Dark Fermion Mass using Multicritical-Point Principle,'' Phys. lett. B 858 139025 (2024), arXiv:2406.07931 [hep-ph].

\bibitem{DM2}G. Arcadi, D. Cabo-Almeida, M. Dutra, P. Ghosh, M. Lindner, Y. Mambrini, J. P. Neto, M. Pierre, S. Profumo, and F. S. Queiroz, ``The Waning of the WIMP:
Endgame?,'' Eur. Phys. J. C 85 152 (2025), arXiv:2403.15860 [hep-ph].

\bibitem{2hdm}L. Wang, J. M. Yang, and Y. Zhang, ``Two-Higgs-doublet models in light of current experiments: a brief review,'' Commun. Theor. Phys. 74 097202 (2022), arXiv:2203.07244 [hep-ph].

\bibitem{higgs3}G. Degrassi, S. Di Vita, J. Elias-Miro, J. R. Espinosa, G. F. Giudice, G. Isidori, and A. Strumia, ``Higgs mass and vacuum stability in the Standard Model at NNLO,'' JHEP 08 098 (2012), arXiv:1205.6497 [hep-ph].


\bibitem{cern}D. Buttazzo et al., ``Investigating the near-criticality of the Higgs boson,'' JHEP 12, 089 (2013), arXiv:1307.3536 [hep-th].

\bibitem{PDG}Particle Data Group et al. , ``Review of Particle Physics,'' PTEP, 083C01 (2022).
\url{https://academic.oup.com/ptep/article/2022/8/083C01/6651666}


\bibitem{bound}Y. Sumino and H. Yokoya, ``Bound-state effects on kinematical distributions of top quarks at hadron colliders,'' JHEP 10, 034 (2010), arXiv:1007.0075 [hep-ph].


\bibitem{bound_ef}B. Fuks, ``Toponium physics at the Large Hadron Collider,'' (2025), arXiv:2505.03869 [hep-ph].

\bibitem{Obs4}M. Misiak and M. Steinhauser, ``Weak Radiative Decays of the $B$ Meson and Bounds on $M_\pm$ in the Two-Higgs-Doublet Model,'' Eur. Phys. J. C 77 201 (2017), arXiv:1702.04571 [hep-ph].
\bibitem{Obs5}A. Arbey, F. Mahmoudi, O. St\r{a}l, and T. Stefaniak, ``Status of the Charged Higgs Boson in Two Higgs Doublet Models,'' Eur. Phys. J. C 78, 182 (2018), arXiv:1706.07414 [hep-ph].

\bibitem{fu1}A. Djouadi, J. Ellis, and J. Quevillon, ``Interference Effects in the Decays of Spin-Zero Resonances into $\gamma\gamma$ and $t\bar{t}$,'' JHEP 07 105  (2016), arXiv:1605.00542 [hep-ph].

\bibitem{fu2}B. Hespel, F. Maltoni, and E. Vryonidou, ``Signal background interference effects in heavy scalar production and decay to a top-anti-top pair,'' JHEP 10 016  (2016), arXiv:1606.04149 [hep-ph].

\bibitem{fu3}P. M. Stephen, ``Two-loop effective potential for a general renormalizable theory and softly broken supersymmetry,'' 	Phys. Rev. D 65 116003 (2002), [hep-ph/0111209].

\bibitem{fu4}A. Anders, F. William, and D. S. Matthew, ``Consistent Use of the Standard Model Effective Potential,'' Phys. Rev. Lett. 113 241801 (2014), arXiv:1408.0292 [hep-ph].

\bibitem{fu5}W. Grimus, L. Lavoura, O.M. Ogreid, and P. Osland, ``The oblique parameters in multi-Higgs-doublet models,'' Nucl. Phys .B 801 (2008), arXiv:0802.4353 [hep-ph].

\end{thebibliography}
\end{document}